\documentclass[aps,twocolumn,amsmath,amssymb,nofootinbib,tighten, superscriptaddress, showpacs ,floatfix]{revtex4-1}

\usepackage{latexsym}
\usepackage{graphicx}
\usepackage{times,psfrag,subfigure}
\usepackage{amsmath}
\usepackage{dcolumn}
\usepackage{bm, bbm}       
\usepackage{color}
\usepackage{latexsym,amsmath,amssymb,bm,euscript}
\def\re#1{(\ref{#1})}
\def\Tr{\mathop{\mathrm{Tr}}}

\newcommand{\beq}{\begin{equation}}
\newcommand{\eeq}{\end{equation}}
\newcommand{\beqarray}{\begin{eqnarray}}
\newcommand{\eeqarray}{\end{eqnarray}}

\begin{document}

 \title{Topological phases and surface flat bands in superconductors without inversion symmetry}

\date{\today}

\author{Andreas P. Schnyder}
\affiliation{Max-Planck-Institut f\"ur Festk\"orperforschung, Heisenbergstrasse 1, D-70569 Stuttgart, Germany}

\author{Shinsei Ryu}
\affiliation{Department of Physics, University of California, Berkeley, CA 94720, USA}

\begin{abstract}
We examine different topological phases 
in three-dimensional non-centrosymmetric superconductors with time-reversal symmetry
by using three different types of topological invariants.
Due to the bulk boundary correspondence, a non-zero value of any of these topological numbers indicates the appearance of zero-energy Andreev surface states. 
We find that some of these boundary modes 
in nodal superconducting phases
are dispersionless, i.e., they form a topologically protected flat band.
The region where the zero-energy flat band appears in the surface Brillouin zone is determined by the projection of the nodal lines in the bulk onto the surface.
These dispersionless Andreev surface bound states have many observable consequences,
in particular, a zero-bias conductance peak
in tunneling measurements.
We also find that in the gapless phase
there appear Majorana surface modes at time-reversal invariant momenta
which are protected by a $\mathbb{Z}_2$ topological invariant.

\end{abstract}

\date{\today}


\maketitle

The hallmark of topological insulators and superconductors (SCs)
is the existence of topologically protected conducting boundary modes.
The recent experimental observation of  these edge and surface states in spin-orbit induced $\mathbb{Z}_2$ topological insulators
in two and three dimensions \cite{konig07ANDhsiehNature08}, respectively, has lead to a surge of interest and 
excitement \cite{Reviews10}.
An exhaustive classification of topologically protected boundary modes occurring in gapped free fermion systems 
in terms of symmetry and spatial dimension
was given in Refs.~\cite{schnyderPRB08, kitaev09, ryuNJP10}. 
Interestingly, this classification scheme, which is known as
the ``periodic table'' of topological insulators and SCs, predicts
a three dimensional (3D) topological SC
which satisfies time-reversal symmetry, but breaks spin-rotation symmetry.
Indeed, the B phase of ${}^3$He is one example of  this so-called ``class DIII'' topological 
superfluid,
whose different
topological sectors can be distinguished by an integer topological invariant.
Recent transverse acoustic impedance measurements in 
${}^3$He-B confirmed the existence
of the predicted 
surface Majorana bound state \cite{murakawa09}.

However, finding an electronic analog of the superfluid B phase of ${}^3$He
remains an outstanding challenge.
In this paper we argue  that some of the 
3D non-centrosymmetric SCs
might be examples of 
electronic topological SCs
in symmetry class DIII  \cite{FootnoteCentroSC}.
We analyze the topological phase diagram of these systems and
demonstrate quite generally
that adjacent to fully gapped topological phases there exist
non-trivial gapless superconducting phases with topologically protected 
nodal lines (rings).
To characterize these gapless lines we 
introduce a set of topological invariants and show that, due to the bulk-boundary correspondence,
the presence of topologically stable nodal rings implies the appearance of 
dispersionless zero-energy Andreev surface states.
These topologically protected  surface flat bands manifest themselves in scanning tunneling spectroscopy (STS) as
a zero bias conductance peak, a feature which could be used 
as an experimental signature of the topological non-triviality.

In non-centrosymmetric SCs the absence of inversion in the crystal structure generates antisymmetric 
spin-orbit couplings (SOC)
and leads to a mixing of spin-singlet and spin-triplet pairing states. These are the properties that give rise to topologically non-trivial quasi-particle band structures in these systems.
Starting with CePt$_3$Si~\cite{bauer04}, a multitude of non-centrosymmetric SCs has recently been discovered, including, 
among others,
Li$_2$Pd$_x$Pt$_{3-x}$B 
\cite{togano04ANDbadica05}.

\emph{Model Hamiltonian}
As a generic phenomenological description applicable to any of the aforementioned materials
we  employ a single band model with antisymmetric 
SOC and 
treat superconductivity at the mean field level.
Thus, let us consider
$
\mathcal{H}
= \sum_{\bm{k}} \Psi^{\dag}_{\bm{k}} H  ( \bm{k} ) \Psi^{\ }_{\bm{k}}$ 
with $\Psi_{\bm{k}} = ( c_{\bm{k} \uparrow  } , c_{\bm{k} \downarrow } , c^{\dag}_{- \bm{k} \uparrow } , c^{\dag}_{- \bm{k} \downarrow } )^T $, where
$c^{\dag}_{\sigma \bm{k}}$ is 
the electron creation operator with spin $\sigma$ and momentum $\bm{k}$ and
the Bogoliubov-de Gennes  (BdG) Hamiltonian is given by  
\begin{eqnarray} \label{ham1a} \label{defHam}
H  ( \bm{k} )
&=&
\begin{pmatrix}
h ( \bm{k} ) & \Delta ( \bm{k} ) \cr
\Delta^{\dag} ( \bm{k} ) & - h^T ( - \bm{k} )
\end{pmatrix} .
\end{eqnarray}
The normal state Hamiltonian $h(\bm{k})$ describes non-interacting electrons
in a crystal without inversion center
$h ( {\bm{k}} )=
\varepsilon_{\bm{k}} \sigma_0 + \bm{\gamma}_{\bm{k}} \cdot \bm{\sigma} ,$
 where $\varepsilon_{ \bm{k} } = \varepsilon_{- \bm{k} }$ is the spin-independent part of the spectrum,
 $\sigma_{1,2,3}$ stand for the three Pauli matrices, and $\sigma_0$ denotes the $2 \times 2$ unit matrix. 
The second term in 
$h ( {\bm{k}} )$
represents an antisymmetric SOC
with coupling constant $\bm{\gamma}_{\bm{k}}$.

Due to the presence of the parity breaking 
SOC  $\bm{\gamma}_{\bm{k}}$ 
the order parameter
in Eq.~\eqref{ham1a} 
is in general an admixture of spin-singlet $\psi_{\bm{k}}$ and spin-triplet $\bm{d}_{\bm{k}}$ pairing states
$\Delta ( {\bm{k}} )
=
\left( \psi_{\bm{k}} \sigma_0 + \bm{d}_{\bm{k} } \cdot \bm{\sigma} \right)
\left( i \sigma_2 \right) ,$
where
$\psi_{\bm{k}}$ and $\bm{d}_{\bm{k}}$ are
even and odd functions of $\bm{k}$, respectively.
The direction of the spin-triplet component $\bm{d}_{\bm{k}}$ is assumed to be parallel to $\bm{\gamma}_{\bm{k}}$, as for this choice the antisymmetric 
SOC is not destructive for triplet pairing \cite{frigeri04}. Hence, we parametrize the $\bm{d}$-vector and the 
SOC as $\bm{d}_{\bm{k}} = \Delta_t \bm{l}_{\bm{k}}$
and $\bm{\gamma}_{\bm{k}} = \alpha \bm{l}_{\bm{k}}$,
respectively.  For the spin-singlet component we assume $s$-wave pairing
$\psi_{\bm{k}} = \Delta_s$ and choose the amplitudes $\Delta_{t,s}$ to be real and positive.

In order to exemplify the topological properties of the BdG Hamiltonian \eqref{defHam}, we consider a
normal state tight-binding band structure on the cubic lattice
$\varepsilon_{ \bm{k} }
=
t_1 \left( \cos k_x + \cos k_y + \cos k_z \right) - \mu ,
$
with hopping amplitude
$t_1$ and chemical potential $\mu$. 
We will set $(t_1, \mu, \alpha, \Delta_t) = (4.0, 4.8, 1.0, 1.0)$ henceforth.
The specific form of the 
SOC $\bm{\gamma}_{\bm{k}}$ depends on the
crystal structure \cite{samokhin09}, i.e.,  $g \bm{\gamma}_{g^{-1} \bm{k}}  = \bm{\gamma}_{ \bm{k}}$, where $g$
is any symmetry operation in the point group $\mathcal{G}$ of the crystal. 
Having in mind Li$_2$Pd$_x$Pt$_{3-x}$B, we assume for the pseudovector $\bm{l}_{\bm{k}}$ the following
 form compatible with the symmetry requirements
of the cubic point group $O$
\begin{eqnarray} \label{lFO}
\bm{l}_{\bm{k}}
=
 \begin{pmatrix}
\sin k_x \cr \sin k_y \cr \sin k_z
\end{pmatrix}
-
g_2
 \begin{pmatrix}
\sin k_x ( \cos k_y + \cos k_z ) \cr \sin k_y  ( \cos k_x + \cos k_z ) \cr \sin k_z ( \cos k_x + \cos k_y ) 
\end{pmatrix} ,
\end{eqnarray}
with the constant  $g_2$, and where we neglect higher order terms.
Furthermore, we also consider the point group $C_{4v}$, relevant for CePt$_3$Si,
in which case $\bm{l}_{\bm{k}}$ reads
\begin{eqnarray}
\label{cubic SO}
\bm{l}_{\bm{k}}
&=&
 \left(  \sin k_y \hat{\bm{e}}_x - \sin k_x \hat{\bm{e}}_y  \right)
\\
&& \quad
+
g_2 \sin k_x   \sin k_y \sin k_z \left( \cos k_x - \cos k_y \right)   \hat{\bm{e}}_z .
\nonumber
\end{eqnarray}
It is important to note that the quasi-particle band topology of $H (\bm{k})$, as defined by Eq.\ (\ref{defHam}), is mainly determined by the momentum
dependence of $\bm{l}_{\bm{k}}$ along the Fermi surface sheets. Hence,  the results we obtain are expected to remain 
qualitatively unchanged upon inclusion of further-neighbor hopping terms in the band structure $\varepsilon_{\bm{k}}$.

\begin{figure}[t]
\begin{center} 
\includegraphics[width=.45\textwidth,angle=-0]{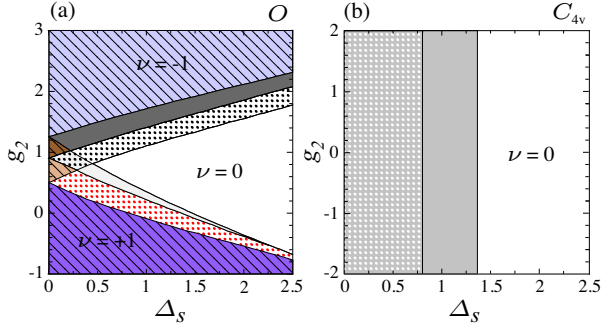}
\caption{
\label{fig:phaseDiags}
(color online).
Phase diagram 
as a function of singlet pairing amplitude $\Delta_s$ and
SOC $g_2$ [see Eqs.~\eqref{lFO} and \eqref{cubic SO}]
for the point group (a) $O$ and (b) $C_{4v}$.
The gapped phases are characterized by the winding number $\nu$ 
with $\nu=0$ (white), $\nu=\pm 1$ (dark/light blue hatched), $\nu=-5$ (light brown hatched), 
and $\nu=+7$ (dark brown hatched).
Grey shaded and dotted regions are nodal superconducting phases with $N_{\mathcal{C}_1}= \pm 1$ (red/black dotted), 
$N_{\mathcal{C}_2}= \pm 1$ (light/dark grey), $N_{\mathcal{C}_3}= + 1$ (white dotted), and $N_{\mathcal{C}_4}= + 1$ (grey).
}
\end{center}
\end{figure}

\emph{Topological Invariants} 
To characterize the topological properties of $H ( \bm{k} )$ we introduce three different topological invariants. 
But before doing so, we observe that $H (\bm{k} )$ satisfies both time-reversal symmetry (TRS), with $\mathcal{T}^2 = -1$, and
particle-hole symmetry (PHS), with $\mathcal{C}^2=+1$, which are the defining symmetry properties of 
symmetry class DIII in the terminology of Ref.~\cite{schnyderPRB08}. 
Combining TRS and PHS
yields a third discrete symmetry, the ``chiral'' symmetry $\mathcal{S} = \mathcal{T} \mathcal{C}$, i.e., there is a unitary matrix
$\mathcal{S}$ which anticommutes with $H ( \bm{k} )$. It is important to note that while
both TRS and PHS relate $H (\bm{k} )$ to $H^T (- \bm{k} )$,  $\mathcal{S}$ is a symmetry
which is satisfied by $H ( \bm{k} )$ at any given point $\bm{k}$ in the Brillouin zone (BZ).

As shown in Ref.~\cite{schnyderPRB08}
topological sectors in the fully gapped phases of $H(\bm{k})$
are distinguished by 
the winding number
\begin{eqnarray} \label{Wno}
\nu 
=
\int_{\mathrm{BZ}} \frac{d^3 k}{24\pi^2} \varepsilon^{\mu \nu \rho}
\Tr \left[ 
( q^{-1} \partial_\mu q ) ( q^{-1} \partial_\nu q ) ( q^{-1} \partial_\rho q )
\right],
\end{eqnarray}
where the integral is over the 1st 
BZ and 
$q ( \bm{k} )$ is the off-diagonal block of the flat-band matrix of $H ( \bm{k} )$~\cite{supplement}. 

In the nodal superconducting phases the winding number $\nu$ is no longer quantized. However, 
we can consider $H ( \bm{k} )$ restricted to 1D loops in reciprocal space and define a  topological number
in terms of a 1D momentum space loop integral to characterize the topology of the gapless phases. 
We observe that $H ( \bm{k} )$ confined to a generic momentum space
loop no longer satisfies TRS nor PHS, but it still obeys chiral symmetry $\mathcal{S}$. 
Hence, $H(\bm{k})$ restricted to a loop in the BZ belongs to symmetry class AIII~\cite{schnyderPRB08}
and its topological characteristics are described by
the 1D winding number
\begin{eqnarray} \label{1Dwind}
N_{\mathcal{L}}
=
\frac{1}{2 \pi i} 
\oint_{\mathcal{L}} d l  \, \mathrm{Tr} \, 
[
q^{-1}(\bm{k}) \nabla_l q (\bm{k}) 
] ,
\end{eqnarray}
where the integral is evaluated along the loop $\mathcal{L}$ in the BZ.
 Observe that for  {\it any} closed loop $\mathcal{L}$ 
that does not intersect with gapless regions in the BZ, $N_{\mathcal{L}}$ is quantized to integer values.
If  $\mathcal{L}$ is chosen such that it  encircles a line node,
then $N_{\mathcal{L}}$ determines the topological stability 
(i.e., the topological charge) of the gapless line \cite{beriPRB2010,sato06}.

Finally, we 
also consider $H ( \bm{k})$ restricted to  a
time-reversal invariant (TRI) loop $\mathcal{L}$, which is mapped onto itself under
${\bm k}\to -{\bm k}$.
In that case we obtain a 
1D Hamiltonian satisfying 
both TRS and PHS (i.e., belonging to symmetry class DIII). The topological properties
of such a 1D
system are characterized by the following  $\mathbbm{Z}_2$ invariant
\cite{supplement}
\begin{eqnarray} \label{Z2No}
W_{\mathcal{L}}
=
\prod_{\boldsymbol{K}} 
 \mathrm{Pf} \left[ q^T (\boldsymbol{K})  \right] /
 \sqrt{ \det \left[ q ( \boldsymbol{K} ) \right] }
,
\end{eqnarray}
where $\boldsymbol{K}$ denotes the two 
TRI momenta on the loop $\mathcal{L}$ and $\mathrm{Pf}$ is the Pfaffian.
Note that $W_{\mathcal{L}}$ is either +1 or -1 
for any TRI loop that does not cross gapless regions in the BZ.

\begin{figure}[t]
\begin{center} 
\vspace{-0.2cm}
\hspace{-0.2cm}
\includegraphics[width=0.45\textwidth,angle=-0]{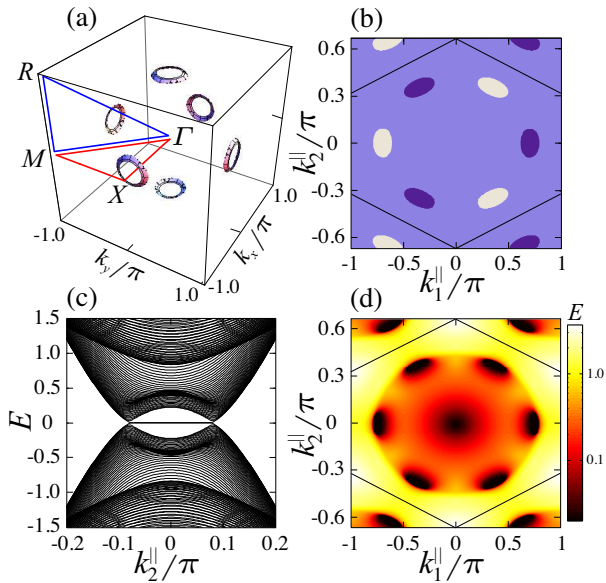}
\caption{
\label{fig:Li2PdPtB}
(color online)
Nodal rings (a) and (111) surface states (c,d) 
for the point group $O$ with
$( g_2, \Delta_s)=( 0.3,0.5)$.
This parameter choice corresponds to the red dotted region in Fig.~\ref{fig:phaseDiags}a.
(b) Topological invariant $N_{(111)}$, Eq.~\eqref{winding1d111}, 
as a function of surface momentum $\bm{k}_{\parallel}$.
Grey and dark blue indicate $N_{(111)} = \pm 1$, while 
light blue is $N_{(111)} =0$.
(c) Band structure for a slab with (111) face as a function of surface momentum $k_2^{\parallel}$ with $k^{\parallel}_1= 0.75 \pi$.
(d) Energy dispersion of  the lowest lying state with positive energy. The color scale is such that black corresponds to zero energy.
The states at zero energy in (c) and (d) are localized at the surface.
The flat bands in (c) and (d) are singly degenerate (i.e., one branch per surface), whereas the 
linearly dispersing zero mode at the center of the BZ in (d)
is doubly degenerate.
}
\end{center}
\end{figure}

\begin{figure}[t]
\begin{center} 
\vspace{-0.2cm}
\hspace{-0.2cm}
\includegraphics[width=.46\textwidth]{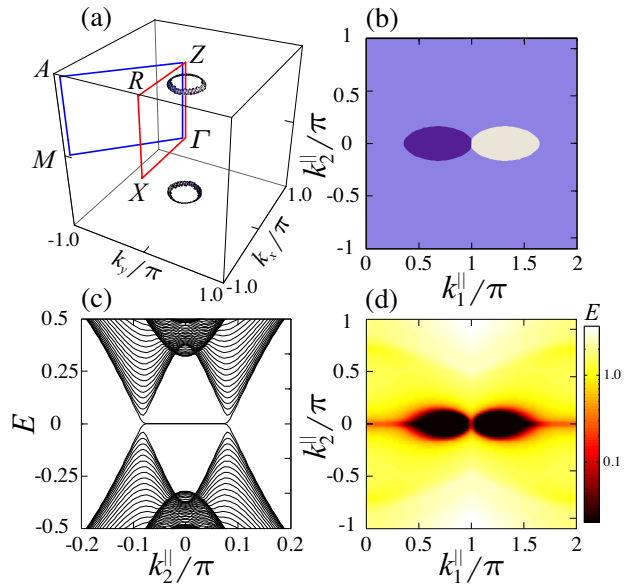}
\caption{
\label{fig:CePt3Si}
(color online) 
Same as Fig.~\ref{fig:Li2PdPtB} but for
the point group $C_{4v}$,  for a slab with (012) face, and with
$( g_2, \Delta_s)=(0.0,0.5)$. This
parameter choice corresponds to the white dotted area in Fig.~\ref{fig:phaseDiags}b.
}
\end{center}
\end{figure}

\emph{Topological Phase diagram}  
Numerical evaluation of  the topological numbers \eqref{Wno} and \eqref{1Dwind} 
yields the topological phase diagram of $H ( \bm{k} )$, 
which is shown in Fig.~\ref{fig:phaseDiags}
as a function of second order
SOC $g_2$ and relative strength of singlet and triplet pairing components.
Fully gapped phases with different topological properties 
(i.e., the phases labeled by $\nu=\pm 1, 0, -5, +7$) 
are separated in the phase diagram by regions 
of nodal superconducting 
phases (grey shaded and dotted areas).
The fully gapped phases with $\nu= \pm 1$ are electronic analogs
of ${}^3$He-B. The nodal superconducting phases
exhibit topologically stable nodal rings,
which are centered around high symmetry axes of the BZ (see Figs.~\ref{fig:Li2PdPtB}a and~\ref{fig:CePt3Si}a).
In order to determine the topological character of these nodal lines (and hence of the corresponding gapless phases) it is sufficient to consider
the topological invariant $N_{\mathcal{L}}$ only for loops $\mathcal{L}$ that run along high symmetry axes.
Thus, for the cubic point group $O$ we choose the loops  
$
\mathcal{C}_1 : \Gamma \to M \to X \to \Gamma
$
and
$
\mathcal{C}_2 : \Gamma \to M \to R \to \Gamma
$,
whereas for the tetragonal point group $C_{4v}$ we consider
$
\mathcal{C}_3:
\Gamma  \to Z \to R \to X \to \Gamma
$
and
$
\mathcal{C}_4:
\Gamma \to Z \to A \to M \to \Gamma
$.
For the cubic point group we find that whenever $( N_{\mathcal{C}_1}, N_{\mathcal{C}_2} )  = ( \pm 1, 0)$ 
there are topologically stable nodal rings centered around the (100) axis (and symmetry related directions).
When $( N_{\mathcal{C}_1}, N_{\mathcal{C}_2} ) = (0 ,  \pm 1)$ the gapless lines are oriented along
 the (111) axis, whereas when $( N_{\mathcal{C}_1}, N_{\mathcal{C}_2} ) = (\pm 1 ,  \pm 1)$
the rings are located around the (110) direction.
(A similar analysis also holds for the group $C_{4v}$.)

\emph{Andreev surface states}
A non-zero quantized value of any of the three topological numbers \eqref{Wno}, \eqref{1Dwind} and \eqref{Z2No} implies the 
existence of zero-energy Andreev surface states. First of all, in fully gapped phases with
topologically non-trivial character there appear linearly dispersing Majorana surface modes 
\cite{schnyderPRB08,qiHughesRaghuZhangPRL09,vorontsov08,eschrig10}.
In order to understand the appearance of zero-energy Andreev surface states 
in the gapless phases, 
we now make use of the topological invariant 
$N_{\mathcal{L}}$ with a cleverly chosen loop  $\mathcal{L}$.
Let us consider
Eq.~\eqref{defHam} in a slab configuration with $(lmn)$ face.
In this geometry the
 Hamiltonian $H_{(lmn)}$
retains translational invariance along the two independent directions 
parallel to the $(lmn)$ surface.
Hence, $H_{(lmn)} (\bm{k}_{\parallel})$ 
can be viewed as a family of 
1D systems parametrized
by the two surface momenta $\bm{k}_{\parallel} = ( k^{\parallel}_1, k^{\parallel}_2 )$.
Since $H_{(lmn)} ( \bm{k}_{\parallel})$ obeys chiral symmetry 
(but breaks in general TRS and PHS),
its topological properties are given by the 
1D winding number of class AIII
 \begin{eqnarray}
 \label{winding1d111}
 N_{(lmn)} (\bm{k}_{\parallel} )
 =
 \frac{1}{2 \pi i} \int d k_{\perp} 
  \, \mathrm{Tr} 
 \left[
 q^{-1}(\bm{k}) 
\partial_{\perp}
q(\bm{k}) 
 \right],
 \end{eqnarray}
where $k_{\perp}$ is the bulk momentum perpendicular to the surface, 
and 
$\partial_{\perp}=\partial/\partial k_{\perp}$.
Note that $N_{(lmn)}$ is the same as $N_{\mathcal{L}}$, Eq.~\re{1Dwind}, with
$\mathcal{L}$ chosen 
along  $k_{\perp}$, 
following a non-contractible
cycle of the BZ torus $T^3$. 

Now, the key observation is that the above line integral 
is closely related the loop integral $N_{\mathcal{L}}$, with $\mathcal{L} = \mathcal{C}_i$,  
that determines the topological charge of the superconducting nodal lines.  
That is, for those surface momenta $\bm{k}_\parallel$ for which the loop along $k_\perp$ in Eq.~\re{1Dwind} passes through
just one non-trivial nodal ring,  $N_{(lmn)} (\bm{k}_{\parallel} )$ is equal to the topological charge of this given nodal ring. 
Hence, if we plot 
$N_{(lmn)} (\bm{k}_{\parallel} )$
as a function of surface momenta 
(see Figs.~\ref{fig:Li2PdPtB}b,\ref{fig:CePt3Si}b), we find
that the boundaries separating regions with different winding number are identical
to the projection of the nodal lines onto the 
$(lmn)$ plane. 
Furthermore, since
a non-zero quantized value of $N_{(lmn)}$ implies the existence
of zero energy states at the end points of the 1D  Hamiltonian  $H_{(lmn)} ( \bm{k}_{\parallel})$
\cite{ryu2002, schnyderPRB08}, 
we find that there are zero-energy Andreev bound states on the $(lmn)$ surface 
located
within the projected nodal rings. This conclusion is corroborated by numerical computations
of the zero-energy surface states both for the point group $O$ and
$C_{4v}$ (see Figs.~\ref{fig:Li2PdPtB} and \ref{fig:CePt3Si}).
When two nodal rings overlap in the $(lmn)$ projection of the BZ, then the quantized 
value of $N_{(lmn)}$ in the overlapping region is determined by the additive contribution
of the topological charges of the two rings. In particular, one can have a situation where the
two contributions cancel, in which case there is no zero-energy surface state within the overlapping region.

Finally, using an analogous argument as in the previous paragraph, we can also employ 
 the $\mathbb{Z}_2$ number \re{Z2No} to deduce the presence of zero energy modes at TRI momenta of the surface BZ \cite{supplement}.
One example of this is the Kramers pair of surface zero modes located at
the center of the surface BZ in Fig.\ \ref{fig:Li2PdPtB}d (cf.\ Refs.~\cite{vorontsov08,eschrig10}).
Remarkably, this is a surface Majorana mode in a {\it gapless} (nodal) superconducting phase \cite{SatoFujimoto2010}.

\begin{figure}[t]
\begin{center} 
\includegraphics[width=.43\textwidth,angle=-0]{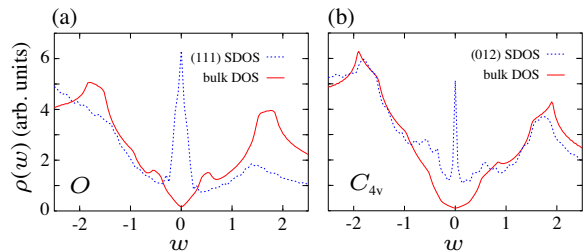}
\caption{
\label{fig:SDOS}
(color online).
Surface and bulk density of states 
for the point group (a) $O$ and (b) $C_{4v}$. The surfaces are oriented
perpendicular to the (111) and (012) axes, respectively.
The employed parameter values are the same as in Figs.~\ref{fig:Li2PdPtB} and \ref{fig:CePt3Si}.
}
\end{center}
\vspace{-0.4cm}
\end{figure}

\emph{Experimental signatures}
One of the most direct signatures of the 
topological aspects
of non-centrosymmetric SCs
are the surface Andreev bound states.
These can be probed by angle-resolved
photoemission measurements, or by 
STS of the surface
density of states (SDOS).  
STS has proved to be an effective tool to explore 
surface states
 of two-dimensional unconventional superconductors,
see, e.g., \cite{tanakaKashiPRL,kashiwayaRepProg10,tanakaPRL10, yadaPRB11,yadaLong}.
The bulk density of states of 3D gapless SCs  with nodal lines vanishes linearly at zero energy. 
In contrast, the surface flat bands lead to a diverging zero-energy peak in the SDOS (see Fig.~\ref{fig:SDOS}).

The zero-bias peak in the SDOS is strongly dependent on the surface orientation.
From this dependence it is in principle possible to (partially) map out 
the location of the topologically stable nodal lines in the bulk BZ.
In addition, one can take advantage of the fact that an applied magnetic field 
leads to a splitting of the zero-energy peak. 
Again, this splitting is strongly dependent on the orientation of the magnetic 
field axis with respect to the nodal lines.
Another possibility is to use spatially resolved STS to investigate the SDOS 
in the presence of impurities on the surface.
It is expected that surface impurities will lead to strong spatial modulations 
of the SDOS, which might give some 
information about the topological characteristics of the nodal lines in the bulk.

\emph{Stability of surface modes}
The zero-energy surface flat bands in 
time-reversal symmetric  
non-centrosymmetric SCs
are topologically protected against the opening of a gap
and are therefore stable against weak symmetry preserving
deformations. 
Conversely,
any perturbation 
that leads to a gap opening of the surface states 
is expected to be accompanied by 
the breaking of 
the symmetries of the 
time-reversal symmetric 
SC, 
i.e., TRS or certain types of translational invariance.
One possible scenario, for example,
is that interactions might lead to spontaneous TRS breaking at the boundary of the SC, 
such as
to the coexistence of TRS breaking 
and TRS preserving order parameters near the surface.
This would be observable in experiments, for instance, as a splitting of the zero-bias conductance peak.

In conclusion, using three different topological invariants, 
we examined the topological properties of general 3D 
non-centrosymmetric superconductors with TRS.
We showed that in nodal superconducting phases there
always appear dispersionless Andreev surface bands.
We established a correspondence between these zero-energy surface flat bands and
the topologically protected nodal lines in the bulk, thereby revealing the topological origin
of the surface flat band. In particular, we demonstrated that the projection of the nodal 
lines on the surface coincides with the boundary of the surface flat band.
We emphasize that the presented formalism (or a generalization thereof) can be
applied to any 3D unconventional SC that preserves TRS. 
One particularly interesting family of compounds is Li$_2$Pd$_x$Pt$_{3-x}$B.
In these SCs the substitution of Pd by Pt seems to be related
to the relative strength of singlet and triplet pairing states 
\cite{yuan06}.
Hence, it might be possible to observe in Li$_2$Pd$_x$Pt$_{3-x}$B  the transition between two
topologically  distinct quantum phases as a function of Pt concentration.

\emph{Acknowledgments}
The authors thank 
B.\ B\'eri, 
A.\ Furusaki,
L.\ Klam, 
A.\ Ludwig, 
R.\ Nakai, 
P.\ Horsch, 
and M.\ Sigrist 
for discussions.
A.P.S. is grateful to the Aspen Center for Physics for hospitality during the preparation of this work.
S.R. is supported by Center for Condensed Matter Theory
at UCB.


\vspace{0.7cm}

 \begin{center}
\textbf{
\Large{Supplementary Materials}
}
\end{center}

 \appendix

\vspace{0.5cm}

We first discuss basic symmetry properties of  superconductors with
time-reversal invariance and then go on to derive the topological 
numbers \eqref{Wno}, \eqref{1Dwind}, and 
\eqref{Z2No} from the main text. 
We shall keep the analysis as general as possible, such that it may be applied to arbitrary  
superconducting systems. In Section \ref{specNCS} we will then specialize to the Bogoliubov-de Gennes Hamiltonian \eqref{defHam}  describing a single-band non-centrosymmetric superconductor.

\section{Symmetries of the Bogoliubov-de Gennes Hamiltonian}

Let us consider a general time-reversal invariant superconductor 
belonging to symmetry class DIII in the terminology of Refs.~\cite{schnyderPRB08,zirnbauerMathPhys96,altlandZirnbauer97}
\begin{eqnarray} \label{genHam}
H ( \bm{k} )
=
\begin{pmatrix}
h ( \bm{k} ) & \Delta ( \bm{k} ) \cr
\Delta^{\dag} ( \bm{k} ) & - h^T ( - \bm{k} ) \cr
\end{pmatrix} ,
\end{eqnarray}
with the $N$-band normal state Hamiltonian $h ( \bm{k} )$ and the superconducting gap matrix $\Delta ( \bm{k} )$, which
obeyes $\Delta ( \bm{k} ) = - \Delta^T ( - \bm{k} )$ because of Fermi statistics.
With $N$ bands (orbitals) and two spin degrees of freedom,
the total dimension of the Bogoliubov-de Gennes Hamiltonian at momentum $\bm{k}$
is $4N\times 4N$.
A class DIII superconductor satisfies two independent anti-unitary symmetries: time-reversal symmetry $\mathcal{T} = \mathcal{K} U_T$, with $\mathcal{T}^2=-1$, and particle-hole symmetry $\mathcal{C} = \mathcal{K} U_C$,
with $\mathcal{C}^2 = + 1$. Here, $\mathcal{K}$ stands for the complex conjugation operator. 
Time-reversal symmetry constrains $H ( \bm{k} )$ as
\begin{eqnarray} \label{TRSeq}
U_T H^{\ast} ( - \bm{k} ) U^{\dag}_T 
=
+ H ( \bm{k} ) ,
\end{eqnarray}
with $U_T = \mathrm{diag}  ( u_T ,  u_T^{\ast} )$ and $u_T$ is a $2N \times 2N$ unitary matrix that 
implements time-reversal invariance of the normal state Hamiltonian, i.e.,
$u_T h^{\ast} ( - \bm{k} ) u^{\dag}_T = h ( \bm{k} )$ and $u^T_T = - u_T$.
Observe, Eq.~\eqref{TRSeq} implies
$u_T \Delta^{\dag} (  \bm{k} )  =  \Delta ( \bm{k} ) u_T^{\dag} $.
Particle-hole symmetry acts on the Bogoliubov-de Gennes Hamiltonian $H ( \bm{k} )$ as
\begin{eqnarray}
U_C H^{\ast} ( - \bm{k} ) U^{\dag}_C
=
-  H ( \bm{k} ) ,
\end{eqnarray}
where $U_C = \sigma_1 \otimes \mathbbm{1}_{2N}$, $\sigma_{1,2,3}$ stand for the three Pauli matrices, and $\mathbbm{1}_{2N}$ is
the $2N \times 2N$ unit matrix. 
Combining time-reversal and particle-hole symmetry we obtain a third
discrete symmetry, which is given by
\begin{eqnarray}
U_{S}^{\dag} H ( \bm{k} ) U_{S} 
= - H ( \bm{k} ) ,
\end{eqnarray}
with 
$U_{S}= i U_T U_C$. 
In other words, there is a unitary matrix 
$U_{S}$ 
that anticommutes with $H ( \bm{k} )$ and thereby
endows the Hamiltonian with a ``chiral" structure. 
Namely, in the basis in which 
$U_{S}$ 
is diagonal, $H ( \bm{k} )$ takes block off-diagonal form
\begin{eqnarray} \label{blockOffHam}
\tilde{H} ( \bm{k} )
=
V H ( \bm{k} ) V^{\dag}
=
\begin{pmatrix}
0 & D ( \bm{k} ) \cr
D^{\dag} ( \bm{k} ) & 0 \cr
\end{pmatrix} ,
\end{eqnarray}
where the unitary transformation $V$ is given by
\begin{eqnarray} \label{Vtraf}
V =
\frac{1}{\sqrt{2}}
\begin{pmatrix}
\mathbbm{1}_{2N}  & + i u_T \cr
\mathbbm{1}_{2N} & - i u_T
\end{pmatrix}  ,
\end{eqnarray}
and the block off-diagonal component  reads
$
D(\bm{k} ) 
=
h ( \bm{k} ) + i \Delta ( \bm{k} ) u_T^{\dag}  .
$
In the off-diagonal basis the unitary matrix $U_T$ is given
as $\tilde{U}_T = V U_T V^T = \sigma_1 \otimes u_T$.
Thus, time-reversal symmetry acts on $D ( \bm{k} )$ as follows
\begin{eqnarray}
u_T D^T ( - \bm{k} ) u_T^{\dag} 
=
D ( \bm{k} ) .
\end{eqnarray}
To compute the $\mathbbm{Z}_2$ invariant, Eq.~\eqref{Z2No}, 
it is advantageous to perform a second basis transformation which
brings  $U_T$ into the simple form
$\bar{U}_T = W U_T W^T =  i \sigma_2 \otimes \mathbbm{1}_{2N} $. This can be achieved with the help of
the unitary matrix
$
W = 
\mathrm{diag} ( \mathbbm{1}_{2N} ,   u_T ) .
$
In this new basis the Bogoliubov-de Gennes Hamiltonian reads
\begin{eqnarray} \label{bar-basis}
\bar{H} ( \bm{k} )
=
W \tilde{H} ( \bm{k} ) W^{\dag}
=
\begin{pmatrix}
0 & \bar{D} ( \bm{k} ) \cr
\bar{D}^{\dag} ( \bm{k} ) & 0 \cr
\end{pmatrix},
\end{eqnarray}
with the block off-diagonal component
\begin{eqnarray}
\bar{D} ( \bm{k} ) = D ( \bm{k} ) u^{\dag}_T  = h ( \bm{k} ) u^{\dag}_T - i \Delta ( \bm{k} ) .
\end{eqnarray}
We note that time-reversal symmetry operates on $\bar{D} ( \bm{k} )$ as
$\bar{D}^T ( - \bm{k} ) = - \bar{D} ( \bm{k} )$.

\section{Flat Band Hamiltonian and Winding~Number}
\label{windingNo}

For the derivation of the topological invariants it is convenient 
to adiabatically deform $H ( \bm{k})$, Eq.~\eqref{genHam}, into a flat band Hamiltonian $Q (\bm{k} )$.
The only assumptions that we need for computing $Q ( \bm{k} )$ are: (i) the Hamiltonian has a full spectral gap and (ii)
there is a unitary matrix $U_S$
anticommuting  with $H ( \bm{k} )$. Thus, the following
derivation of $Q ( \bm{k} )$ is applicable to any chiral symmetric Hamiltonian with a full bulk gap, in 
particular also to the three-dimensional topological superconductors in 
symmetry class AIII, DIII, and CI 
\cite{schnyderPRB08,ryuNJP10,volovikBOOKS,roy2008,zhangPRB08,schnyderPRL09,schnyderPRB10}.

In what follows, we work in a basis in which $H ( \bm{k} )$ takes block off-diagonal form.  The flat band Hamiltonian is defined in terms of the projection 
operator $P ( \bm{k})$ which projects onto filled Bloch eigenstates of $H (\bm{k})$ at a given momentum $\bm{k}$.
The projector $P ( \bm{k} )$, in turn, is defined in terms of the eigenfunctions of $ \tilde{H} (\bm{k} )$
\begin{eqnarray} \label{eigenEQ}
\begin{pmatrix}
0 & D ( \bm{k} ) \cr
D^{\dag} ( \bm{k} ) & 0 
\end{pmatrix}
\begin{pmatrix}
\chi^{\pm}_a ( \bm{k} ) \cr
\eta^{\pm}_a ( \bm{k} ) \cr
\end{pmatrix}
=
\pm \lambda_a ( \bm{k} ) 
\begin{pmatrix}
\chi^{\pm}_a ( \bm{k} ) \cr
\eta^{\pm}_a ( \bm{k} ) \cr
\end{pmatrix} , \quad
\end{eqnarray}
where 
$a=1,\ldots,2N$
is the combined band and spin index.
We assume there is a spectral gap around zero energy with
 $\left| \lambda_a ( \bm{k} ) \right| > 0$, and for definitiveness we choose $\lambda_a ( \bm{k} ) > 0$ for all $a$.
Multiplying equation \eqref{eigenEQ}  from the left  by $\tilde{H} ( \bm{k} )$  yields 
\begin{eqnarray}
DD^{\dag} \chi^{\pm}_a ( \bm{k} ) 
= \lambda^2_a \chi^{\pm}_a ,
\quad
D^{\dag} D \eta^{\pm}_a 
= 
\lambda^2_a \eta^{\pm}_a . 
\end{eqnarray}
Hence, the eigenfunctions  $( \chi^{\pm}_a, \eta^{\pm}_a)$
can be obtained from the eigenvectors of $DD^{\dag}$ or $D^{\dag} D$
\begin{eqnarray} \label{evalsDD}
D D^{\dag} u_a  = \lambda^2_a u_a , 
\quad
D^{\dag} D v_a  = \lambda^2_a v_a . 
\end{eqnarray}
The eigenvectors $u_a$, $v_a$ are taken to be normalized to one, i.e.,   $u^{\dag}_a u_a = v^{\dag}_a v_a = 1$, for all $a$ (here, the index $a$ is not summed over). 	 
The eigenstates of $D^\dag D$ follow from the eigenstates of $D D^{\dag}$ via 
\begin{eqnarray}
v_a  = \mathcal{N}_a  D^{\dag} u_a,	 
\end{eqnarray}
with the normalization factor $\mathcal{N}_a$.  Using Eq.~\eqref{evalsDD} one can check that $v_a$ is indeed an eigenvector
of $D^{\dag} D$,
\begin{eqnarray}
D^{\dag} D v_a = D^{\dag} D( \mathcal{N}_a D^{\dag} u_a ) = \mathcal{N}_a \lambda^2_a D^{\dag} u_a  = \lambda^2_a v_a ,
\quad
\end{eqnarray}
for all $a$.
The normalization factor $\mathcal{N}_a$ is given by
\begin{eqnarray}
u^{\dag}_a D D^{\dag} u_a = \lambda^2_a u^{\dag}_a  u_a = \lambda^2_a 
\; \Rightarrow  \;  \mathcal{N}_a =  \frac{1}{ \lambda_a } ,
\end{eqnarray}
for all $a$.
It  follows that
the eigenfunctions of $\tilde{H} ( \bm{k} )$ are
\begin{eqnarray}
\begin{pmatrix}
\chi^{\pm }_a \cr
\eta^{\pm}_a
\end{pmatrix}
=
\frac{1}{\sqrt{2} }
\begin{pmatrix}
u_a \cr
\pm v_a
\end{pmatrix}
=
\frac{1}{\sqrt{2}}
\begin{pmatrix}
u_a \cr
\pm D^{\dag} u_a / \lambda_a
\end{pmatrix} .
\end{eqnarray}
With this, the projector $P ( \bm{k} )$  onto the filled Bloch states
becomes
\begin{eqnarray}
P 
&=&
\frac{1}{2}
\sum_a 
\begin{pmatrix}
u_a \cr - v_a 
\end{pmatrix}
\begin{pmatrix}
u^{\dag}_a & - v^{\dag}_a 
\end{pmatrix}
\\
&=&
\frac{1}{2}
\begin{pmatrix}
\mathbbm{1}_{2N} & 0 \cr
0 & \mathbbm{1}_{2N} \cr
\end{pmatrix}
-
\frac{1}{2}
\sum_a 
\begin{pmatrix}
0 &   u_a v^{\dag}_a \cr
  v_a u^{\dag}_a & 0 
\end{pmatrix} .
\nonumber
\end{eqnarray}
Finally, we obtain for the flat band Hamiltonian $Q  $, which 
is defined as $Q  = \mathbbm{1}_{4N} - 2 P$ \cite{schnyderPRB08},
\begin{eqnarray}
Q
= 
\sum_a 
\begin{pmatrix}
0 & u_a v_a^{\dag} \cr
v_a u^{\dag}_a & 0 
\end{pmatrix}
=
\sum_a
\begin{pmatrix}
0 &  u_a u_a^{\dag} \frac{ D }{ \lambda_a } \cr
\frac{ D^{\dag} }{  \lambda_a }   u_a u^{\dag}_a & 0
\end{pmatrix}.
\nonumber\\
\end{eqnarray}
In other words, the off-diagonal block of $Q ( \bm{k} )$ reads
\begin{eqnarray}
q( \bm{k} ) 
=
\sum_a \frac{1}{\lambda_a ( \bm{k} ) }
u_a (\bm{k} ) u^{\dag}_a (\bm{k} ) D( \bm{k}) ,
\end{eqnarray}
where $u_a ( \bm{k} )$ denotes the eigenvectors of $D D^{\dag}$.
For a system with completely degenerate bands,  $\lambda_a =  \lambda $, for all $a$,
the above formula simplifies to
\begin{eqnarray} \label{smallQ}
q (\bm{k} ) =
\frac{1}{ \lambda ( \bm{k} ) }
\sum_a
u_a ( \bm{k} ) u^{\dag}_a ( \bm{k} ) D ( \bm{k} )
=
\frac{1}{ \lambda ( \bm{k} ) } D ( \bm{k} ) .
\quad
\end{eqnarray}
Examples of topological insulators and superconductors with completely degenerate bands
are the Dirac representatives of Ref.~\cite{ryuNJP10}.

The integer-valued topological invariant characterizing
topological superconductors is now simply given
by the winding number of $q ( \bm{k} )$.
It can be defined in any odd spatial dimension.
In three dimensions we have
\begin{eqnarray}
\nu_3 
=
\int_{\mathrm{BZ}} \frac{d^3 k}{24\pi^2} \varepsilon^{\mu \nu \rho}
\mathrm{Tr} \left[ 
( q^{-1} \partial_\mu q ) ( q^{-1} \partial_\nu q ) ( q^{-1} \partial_\rho q )
\right],
\nonumber\\
\end{eqnarray}
and in one spatial dimension it reads
\begin{eqnarray} 
\nu_1
=
\frac{1}{2 \pi i}
\int_{\mathrm{BZ}} d k \,
 \mathrm{Tr} \left[  q^{-1} \partial_k q \right] .
\end{eqnarray}
Alternatively, it is also possible to define the winding number
in terms of the unflattened off-diagonal block $D( \bm{k} )$ of the Hamiltonian.
For example, for the winding number in one spatial dimension this reads
\begin{eqnarray}
\nu_1 
&=&
\frac{1}{4 \pi i}
\int_{\mathrm{BZ}} d k \,
\mathrm{Tr} 
\left[
  D^{-1}
\partial_k  D 
-
\{ D^{\dag}  \}^{-1}
\partial_k
D^{\dag}  
\right]  
\nonumber\\
&=&
\frac{1}{2 \pi }
\mathrm{Im}
\int_{\mathrm{BZ}} d k \,
\mathrm{Tr} 
\left[
\partial_k  
\ln D 
\right] .  
\end{eqnarray}

\section{$\mathbb{Z}_2$ Invariant for Symmetry Class DIII  }
\label{Z2invariant}

In this section we compute 
the $\mathbb{Z}_2$ topological invariant for symmetry class DIII in $d=1$ and $d=2$ spatial dimensions.
It is most convenient to perform this derivation in the basis \eqref{bar-basis},
in which the $4N \times 4N$ Bogoliubov-de Gennes Hamiltonian takes the form
\begin{align}
H (\bm{k} )
&=
\left(
\begin{array}{cc}
0 & D( \bm{k} ) \\
D^{\dag}( \bm{k} ) & 0
\end{array}
\right),
\quad
D( \bm{k} )= -D^T(- \bm{k} ).
\label{diii}
\end{align}
In this representation, the time-reversal symmetry
operator is given by  $\mathcal{T} =  \mathcal{K} U_T =     \mathcal{K} \, {i} \sigma_2 \otimes \mathbbm{1}_{2N} $
and the flat band Hamiltonian reads
\begin{eqnarray} \label{Qzwei}
Q( \bm{k} )
=
\left(
\begin{array}{cc}
0  & q( \bm{k} ) \\
q^{\dag}(\bm{k} ) & 0 
\end{array}
\right),
\quad
q( \bm{k} ) = - q^T (-\bm{k} ). 
\end{eqnarray}
The presence of time-reversal symmetry allows
us to define the Kane-Mele $\mathbb{Z}_2$ invariant \cite{kaneMelea05a, kaneMelea05b,fuKane07, ryuNJP10,moorePRB07,royPRB09}, 
\begin{eqnarray} \label{Z2noS}
W
&=&
 \prod_{\bm{K}}
 \frac{ \mathrm{Pf}\, \left[w(\bm{K} )\right] }
 { \sqrt{ \det \left[  w ( \bm{K} ) \right] }} ,
\end{eqnarray}
with $\bm{K}$ a time-reversal invariant momentum
and $\mathrm{Pf}$ the Pfaffian of an anti-symmetric matrix.
 Here, 
$w ( \bm{k} )$ denotes the ``sewing matrix''
\begin{eqnarray} \label{sewMat}
w_{a b }( \bm{k} ) 
& =& 
\langle u^+_{a} (- \bm{k} ) |\mathcal{T} \,  u^+_{b}( \bm{k} ) \rangle,
\end{eqnarray}
where $a , b = 1, \ldots, 2N $ and
$u^{\pm}_{a}(\bm{k})$ is the $a$-th eigenvector
of $Q (\bm{k} )$ with eigenvalue $\pm 1$.
The Pfaffian is an analog of the determinant that can be defined only for  $2n \times 2n$ anti-symmetric matrices $A$. It is given in terms of a sum
over all elements of the permutation group $S_{2n}$
\begin{eqnarray}
\mathrm{Pf} ( A) 
=
\frac{1}{2^n n! }
\sum_{\sigma \in S_{2n} }
\mathrm{sgn} ( \sigma ) \prod_{i=1}^n A_{\sigma(2i-1), \sigma ( 2 i ) } .
\nonumber
\end{eqnarray}

Due to the block off-diagonal structure of Eq.~\eqref{Qzwei}
a set of eigen Bloch functions of $Q ( \bm{k} )$ can be constructed as \cite{ryuNJP10}
\begin{eqnarray} \label{EVsN}
|u^{\pm}_{{a}}( \bm{k} )\rangle_{\mathrm{N}}
=
\frac{1}{\sqrt{2}}
\left(
\begin{array}{c}
n_{{a}} \\
\pm q^{\dag}( \bm{k} ) n_{{a}}
\end{array}
\right),
\label{u_N}
\end{eqnarray}
or, alternatively, as
\begin{eqnarray}
|u^{\pm }_{{a}}( \bm{k} )\rangle_{\mathrm{S}}
=
\frac{1}{\sqrt{2}}
\left(
\begin{array}{c}
\pm  q( \bm{k} ) n_{{a}} \\
n_{{a}}
\end{array}
\right),
\label{u_S}
\end{eqnarray}
where $n_{{a}}$ are $2 N$ momentum independent orthonormal vectors. For simplicity
we choose $(n_{a})_{{b}}= \delta_{a b}$.
In passing, we note that  both 
$|u^{\pm}_{{a}}(\bm{k})\rangle_{\mathrm{N}}$
and
$|u^{\pm}_{{a}}(\bm{k})\rangle_{\mathrm{S}}$
are well-defined globally over the entire Brillouin zone.
To compute the $\mathbbm{Z}_2$ topological number 
we choose the basis
$|u^{\pm}_{{a}}(\bm{k} )\rangle_{\mathrm{N}}$.
Combining Eqs.~\eqref{sewMat} and \eqref{EVsN} yields
\begin{eqnarray}
w_{{a} {b} } ( \bm{k} ) 
&=&
\frac{1}{2}
\big(
\begin{array}{cc}
n^{\dag}_{a}, & n^{\dag}_{a}q(-\bm{k} )
\end{array}
\big)
{i}\sigma_2 \otimes \mathbbm{1}_{2N } \mathcal{K}
\left(
\begin{array}{c}
n_{b} \\
q^{\dag}( \bm{k} ) n_{b}
\end{array}
\right)
\nonumber \\
&=&
\frac{1}{2} 
\big(
\begin{array}{cc}
n^{\dag}_{a}, & n^{\dag}_{a}q(- \bm{k} )
\end{array}
\big)
\left(
\begin{array}{c}
q^{T}( \bm{k} ) n_{b} \\
-n_{b} 
\end{array}
\right)
\nonumber \\
&=&
\frac{1}{2}
\left(
n^{\dag}_{a}
q^T( \bm{k} ) 
n^{\ }_{ b}
-
n^{\dag}_{ a}
q(- \bm{k} ) 
n^{\ }_{b}
\right)
\nonumber \\
&=&
q^T_{a b}( \bm{k} ) .
\end{eqnarray}
In the second last line we used Eq.~\eqref{Qzwei}, i.e.,
$q(- \bm{k} ) = -q^T( \bm{k})$. 
In conclusion, the $\mathbb{Z}_2$ topological number 
in spatial dimensions $d=2$ and $d=1$
is given by
\begin{eqnarray} \label{eqW}
W   = 
\prod_{\bm{K}} 
\frac{
\mathrm{Pf}\, \left[q^T(\bm{K})\right] 
}{
\sqrt{ \det \left[ q ( \bm{K} ) \right] }} ,
\end{eqnarray}
where $\bm{K}$ denotes 
the four (two) time-reversal invariant momenta 
of the two-dimensional (one-dimensional) Brillouin zone.

\section{Flat Band Hamiltonian
for the Non-centrosymmetric Superconductor, Eq.~(1)}
\label{specNCS}

Let us now apply
the formalism developed in
the preceding sections 
to the Bogoliubov-de Gennes Hamiltonian \eqref{defHam} from the main text, 
describing a single band non-centrosymmetric superconductor \cite{frigeriPRL04,samokhinPRB08}.
First, we note that time-reversal symmetry for $H (\bm{k})$, Eq.~\eqref{defHam}, is implemented
by $U_T H^{\ast} ( - \bm{k} ) U^{\dag}_T = + H ( \bm{k} )$ with $U_T = \sigma_0 \otimes i \sigma_2$. Hence, 
we need to set $u_T = i \sigma_2$ in Eq.~\eqref{TRSeq}.
It then follows from Eq.~\eqref{Vtraf} that $H ( \bm{k} )$ can be brought
into block off-diagonal form by the unitary transformation
\begin{eqnarray}
V
=
\frac{1}{\sqrt{2}}
\left(
\begin{array}{cc}
\mathbbm{1}_2 & - \sigma_2  \\
\mathbbm{1}_2 & + \sigma_2
\end{array}
\right) .
\end{eqnarray}
The transformed Hamiltonian is given by
\begin{eqnarray}
V H ( \bm{k} ) V^{\dag}
=
\left(
\begin{array}{cc}
0 & D( \bm{k} ) \\
D^{\dag}( \bm{k} ) & 0
\end{array}
\right) ,
\end{eqnarray}
with the off-diagonal block  
\begin{eqnarray} \label{offBlockNCS}
D ( \bm{k} )
&=&
\left(
\begin{array}{cc}
 B_{ \bm{k}Ê} + A l^z_{ \bm{k} }
& 
A ( l^x_{\bm{k}} - i l^y_{\bm{k}} )
\\
A  ( l^x_{\bm{k}} + i l^y_{\bm{k}} ) & 
 B_{ \bm{k} } - A l^z_{ \bm{k} } 
\end{array}
\right)
\nonumber \\
&=&
B_{\bm{k}}  \sigma_0 + A \,  \bm{l}_{\bm{k}} \cdot \bm{\sigma} ,
\end{eqnarray}
and where we have introduced  the short-hand notation
\begin{eqnarray}
A = 
\alpha + {i}\Delta_t, 
\quad
B_{\bm{k}} = \varepsilon_{\bm{k}}  + {i} \Delta_s.
\end{eqnarray}
Alternatively, we can also choose to work in the basis \eqref{bar-basis}, 
in which case the off-diagonal component reads
\begin{eqnarray}
\bar{D} ( \bm{k} )
=
\left(
\begin{array}{cc}
A ( l^x_{\bm{k}} - i l^y_{\bm{k}} )
&
 - B_{ \bm{k}Ê} - A l^z_{ \bm{k} }
\\
 B_{ \bm{k} } - A l^z_{ \bm{k} }  &
- A  ( l^x_{\bm{k}} + i l^y_{\bm{k}} ) 
\end{array}
\right) .
\end{eqnarray}
For the computation of the flat band Hamiltonian $Q ( \bm{k} )$ it is, however, 
more convenient to use Eq.~\eqref{offBlockNCS}. 

\begin{figure}[t!]
\begin{center} 
\includegraphics[width=.47\textwidth,angle=-0]{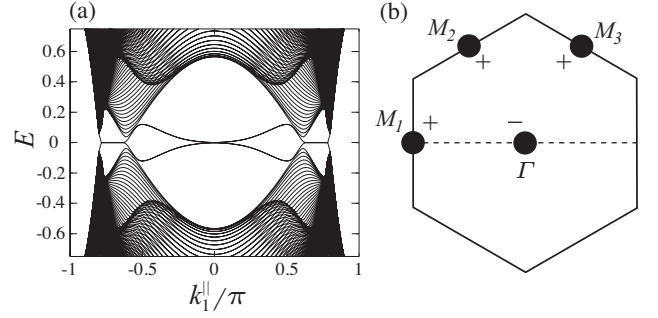}
\caption{
\label{fig:W111}
(a) Band structure of Hamiltonian  \eqref{defHam} for the point group $O$ in a slab geometry with $(111)$ face as a function of surface momentum $k_1^{\parallel}$ with $k^{\parallel}_2= 0$, i.e., along the dashed line in panel (b). Here, we set $( g_2, \Delta_s)=( 0.3,0.5)$.
(b) Brillouin zone of the (111) surface with the values of $W_{(111)} ( \bm{K}_{\parallel} )$, Eq.~\eqref{W2}, at the four time reversal invariant
momenta $\bm{K}_{\parallel} \in \left\{ \Gamma, M_1, M_2, M_3 \right\}$.
}
\end{center}
\end{figure}

Repeating the steps of section \ref{windingNo},
we calcualte the eigenvectors $u_{{a}} ( \bm{k} )$ of
\begin{eqnarray}
D_{\bm{k}} D^{\dag}_{\bm{k}}
=
|A|^2 l^2_{\bm{k}}  + |B_{\bm{k}} |^2 
+
(A B^{\ast}_{\bm{k}}  +B_{\bm{k}} A^{\ast} )
\bm{l}_{\bm{k}} \cdot \bm{\sigma}  , \quad
\end{eqnarray}
where $l_{\bm{k}} = \left|  \boldsymbol{l}_{\bm{k}} \right|$.
The eigenfunctions $u_{{a}} ( \bm{k} )$ of $D^{\ }_{\bm{k}}D^{\dag}_{\bm{k}}$ can be obtained
by diagonalizing $\bm{l}_{\bm{k}} \cdot \bm{\sigma}$ .
Hence, when $(l^x_{\bm{k}} ,l^y_{\bm{k}} )\neq (0,0)$,  we find that the eigenvectors $u_{{a}} ( \bm{k} )$  are given by
\begin{eqnarray}
u_{1 / 2} ( \bm{k} )
&=&
\frac{1}{\sqrt{2  l_{\bm{k}} ( l_{\bm{k}} \mp l^z_{\bm{k}} )}}
\left(
\begin{array}{c}
l^x_{\bm{k}}  -{i} l^y_{\bm{k}}  \\
\pm  l_{\bm{k}} -l^z_{\bm{k}}
\end{array}
\right).
\end{eqnarray}
According to Eq.~\eqref{smallQ},
the off-diagonal block of the flat band Hamiltonian $Q (\bm{k})$ 
is defined in terms of the eigenvectors $u_{{a}} ( \bm{k}) $.
Thus, we need to compute
\begin{eqnarray} \label{interR1}
&&
\sum_{{a} = 1, 2 }
\frac{1}{\lambda_{{a} \bm{k} } }
u^{\ }_{{a}} ( \bm{k} ) u^{\dag}_{{a}} ( \bm{k} )
 \\ 
&=&
\frac{1}{2\lambda_{1\bm{k}} \lambda_{2 \bm{k}} }
\left[
(\lambda_{1 \bm{k}} +\lambda_{ 2 \bm{k}} )\sigma_0
+ 
(\lambda_{2 \bm{k}}-\lambda_{1 \bm{k}} ) 
\frac{\bm{l}_{\bm{k}}  }{ l_{\bm{k}} } \cdot \bm{\sigma}
\right], 
\nonumber
\end{eqnarray}
with the two positive eigenvalues  
$ \lambda_{1 \bm{k}} =  | B_{\bm{k}} -A l_{\bm{k}}   | $ and
$ \lambda_{2 \bm{k}}  = | B_{\bm{k}} + A l_{\bm{k}}    | $.
Note that the last term in the second line of Eq.~\eqref{interR1}
contains  removable singularities at the points $\bm{k}_0$ where
$\bm{l}_{\bm{k}_0} =0$. For those points in the Brillouin zone one needs
to carefully take the limit $\bm{k} \to \bm{k}_0$ to
obtain the correct value of Eq.~\eqref{interR1}.
Finally, by use of Eq.~\eqref{smallQ} together with Eqs.~\eqref{offBlockNCS} and   \eqref{interR1} we find
for the off-diagonal block of the flat band Hamiltonian
\begin{eqnarray}
&& q ( \bm{k} )  
=
\frac{1}{2 \lambda_{1 \bm{k}} \lambda_{2 \bm{k}} }
\Big[
\left\{
A  l_{\bm{k}} (\lambda_{2 \bm{k}} -\lambda_{1 \bm{k}} )
+
B_{\bm{k}}  (\lambda_{1 \bm{k}} + \lambda_{2 \bm{k}} )
\right\}
 \sigma_0
\nonumber \\ 
&&
\qquad
+
\left\{
A l_{\bm{k}} (\lambda_{1 \bm{k}} +\lambda_{2 \bm{k}} )
+
B_{\bm{k}} (\lambda_{2 \bm{k}} -\lambda_{1 \bm{k}} )
\right\}
\frac{\bm{l}_{\bm{k}} }{ l_{\bm{k}} } \cdot \bm{\sigma}
\Big].
\end{eqnarray}

Now, for the $\mathbbm{Z}_2$ invariant we need to bring $q ( \bm{k} )$ into the basis in which $U_T = i \sigma_2 \otimes \mathbbm{1}_2 $. 
This is achieved by letting
\begin{eqnarray}
q ( \bm{k} ) \quad
\to \quad
- i q ( \bm{k} ) \sigma_2  .
\end{eqnarray}
Using Eq.~\eqref{eqW} we get
\begin{eqnarray} \label{WsingelBand}
W
&=&
\prod_{\bm{K} }
\frac{
\mathrm{Pf}\, \left[ i \sigma_2 q^T ( \bm{K} ) \right]
}
{ \sqrt{ \det \left[ i \sigma_2 q^T ( \bm{K} ) \right] }}
= 
\prod_{\bm{K} }
\frac{ B_{\bm{K}}  }
{ \sqrt{ B_{\bm{K}}^2  } } ,
\end{eqnarray}
where we have made use of the fact that $\bm{l}_{\bm{k}}$ is an antisymmetric function, i.e., $\bm{l}_{-\bm{k}} = - \bm{l}_{\bm{k}} $.

\subsection{$\mathbbm{Z}_2$ surface state}

As discussed in the main text, the $\mathbbm{Z}_2$ number  \eqref{WsingelBand} can 
be used to deduce the presence of Andreev surface states at time-reversal invariant
momenta of the surface BZ. To exemplify this, let us consider Hamiltonian \eqref{defHam} in a slab geometry with $(lmn)$ face.
At the four time-reversal invariant momenta $\bm{K}_{\parallel}$ of the $(lmn)$ surface BZ the $\mathbbm{Z}_2$ invariant
is defined by
\begin{eqnarray} \label{W2}
W_{(lmn)} ( \bm{K}_{\parallel} )
&=&
\prod_{\bm{K}_{\perp} }
\frac{
\mathrm{Pf}\, \left[ i \sigma_2 q^T ( \bm{K}_{\perp}, \bm{K}_{\parallel} ) \right]
}
{ \sqrt{ \det \left[ i \sigma_2 q^T ( \bm{K}_{\perp}, \bm{K}_{\parallel} ) \right] }} .
\end{eqnarray}
Eq.~\eqref{W2} is quantized to $+1$ or $-1$, with $W_{(lmn)} ( \bm{K}_{\parallel} ) = -1$ indicating the presence of  Kramers
degenerate surface modes at the surface momentum $\bm{K}_{\parallel}$ (see Fig.~\ref{fig:W111}).


\vspace{-0.2cm}

\begin{thebibliography}{99}

\bibitem{konig07ANDhsiehNature08}
M.\ K\"onig, 
S.\ Wiedmann, C.\ Br\"une, A.\ Roth,
H.\ Buhmann, L.\ W.\ Molenkamp,  X-L.\ Qi, S-C.\ Zhang,
Science \textbf{318}, 766 (2007);
%
D.\ Hsieh, D.\ Qian, L.\ Wray, Y.\ Xia, Y.\ Hor, R.\ Cava, and M.\ Hasan,
Nature \textbf{452}, 970 (2008).

\bibitem{Reviews10}
M.\ Z.\ Hasan and C.\ L.\ Kane,  Rev.\ Mod.\ Phys.\ \textbf{82}, 3045 (2010);
X.-L.\ Qi and S.-C.\ Zhang, arXiv:1008.2026 (unpublished).

\bibitem{schnyderPRB08}
A.\ P.\ Schnyder, S.\ Ryu, A.\ Furusaki, A.\ W.\ W.\ Ludwig, Phys.\ Rev.\ B \textbf{78} 195125 (2008);
AIP Conf.\ Proc.\ \textbf{1134} 10 (2009).

\bibitem{kitaev09}
A.\ Y.\ Kitaev, AIP Conf.\ Proc.\ \textbf{1134} 22 (2009).

\bibitem{ryuNJP10}
S.\ Ryu, A.\ P.\ Schnyder, A.\ Furusaki, A.\ W.\ W.\ Ludwig, New Journal of Physics \textbf{12}  065010 (2010).

\bibitem{murakawa09}
S.\  Murakawa,
Y.\ Tamura, Y.\ Wada, M.\ Wasai, M.\ Saitoh, Y.\ Aoki, R.\ Nomura, Y.\ Okuda, Y.\ Nagato, 
M.\ Yamamoto, S.\ Higashitani, and K.\ Nagai,
Phys.\ Rev.\ Lett.\ \textbf{103}, 155301 (2009).

\bibitem{FootnoteCentroSC}
Topological SCs in class DIII may also be found
among \emph{centrosymmetric} SCs with triplet pairing, see, 
e.g., G.\ E.\ Volovik and L.\ P.\  Gorkov, Sov.\ Phys.\ JETP \textbf{61}, 843 (1985). 

\bibitem{bauer04}

E.\ Bauer, G.\ Hilscher, H.\ Michor, Ch.\ Paul, E.\ W.\ Scheidt, 
A.\ Gribanov, Yu.\ Seropegin, H.\ No\"el, M.\ Sigrist, and P.\ Rogl,
Phys.\ Rev.\ Lett.\ \textbf{92}, 027003 (2004). 

\bibitem{togano04ANDbadica05}
K.\ Togano, P.\ Badica, Y.\ Nakamori, S.\ Orimo, H.\ Takeya, and K.\ Hirata,
Phys.\ Rev.\ Lett.\ \textbf{93}, 247004 (2004);
%
P.\ Badica, T.\ Kondo, and K.\ Togano,
 J.\ Phys.\ Soc.\ Jpn.\ \textbf{74}, 1014 (2005).

\bibitem{frigeri04}
P.\ A.\ Frigeri, D.\ F.\ Agterberg, A.\ Koga, and M.\ Sigrist,
Phys.\ Rev.\ Lett.\ \textbf{92}, 097001 (2004). 

\bibitem{samokhin09}
K.\ V.\ Samokhin,  Annals of Physics \textbf{324}  2385 (2009). 

\bibitem{supplement}
See supplementary material for details of the calculation.

\bibitem{beriPRB2010} 
B.\ B\'eri, Phys.\ Rev.\ B \textbf{81}, 134515 (2010).

\bibitem{sato06}
M.\ Sato, Phys.\ Rev.\ B \textbf{73} 214502 (2006).

\bibitem{qiHughesRaghuZhangPRL09}
X.-L.\ Qi, T.\ L.\ Hughes, S.\ Raghu, and S-C.\ Zhang, 
Phys.\ Rev.\ Lett.\ \textbf{102} 187001 (2009).

\bibitem{vorontsov08}
A.\ B.\ Vorontsov, I.\ Vekhter, and M.\ Eschrig,
Phys.\ Rev.\ Lett.\ \textbf{101}, 127003 (2008).

\bibitem{eschrig10}
M.\ Eschrig, C.\ Iniotakis, and Y.\ Tanaka, arXiv:1001.2486.

\bibitem{ryu2002}
S.\ Ryu and Y.\ Hatsugai, Phys.\ Rev.\ Lett.\ \textbf{89}, 077002 (2002).

\bibitem{SatoFujimoto2010}
For a similar Majorana Andreev state at edges of 2D nodal superconductors, 
see M.\ Sato and S.\ Fujimoto, Phys.\ Rev.\ Lett.\ \textbf{105}, 217001 (2010).

\bibitem{tanakaKashiPRL}
Y. Tanaka and S. Kashiwaya, Phys. Rev. Lett. \textbf{74}, 3451 (1995).

\bibitem{kashiwayaRepProg10}
S.\ Kashiwaya and Y.\ Tanaka, Rep.\ Prog.\ Phys.\ \textbf{63}, 164 (2000).

\bibitem{tanakaPRL10}
 Y.\ Tanaka, Y.\ Mizuno, T.\ Yokoyama, K.\ Yada, and M.\ Sato
 Phys.~Rev.~Lett.\ \textbf{105}, 097002 (2010).

\bibitem{yadaPRB11}
K.\ Yada, M.\ Sato, Y.\ Tanaka, and T.\ Yokoyama
Phys.\ Rev.\ B \textbf{83}, 064505 (2011).

\bibitem{yadaLong}
M.\ Sato, Y.\ Tanaka, K.\ Yada, and T.\ Yokoyama, Phys.\ Rev.\ B \textbf{83}, 224511 (2011).

\bibitem{yuan06} 
H.\ Q.\ Yuan, D.\ F.\ Agterberg, N.\ Hayashi, P.\ Badica, D.\ Vandervelde, K.\ Togano, M.\ Sigrist, and M.\ B.\ Salamon,
Phys.\ Rev.\ Lett.\  \textbf{97}, 017006 (2006).



\end{thebibliography}

\begin{thebibliography}{99}

\bibitem[26]{zirnbauerMathPhys96}
M.\ R.\ Zirnbauer, J. Math. Phys. \textbf{37}, 4986 (1996). 

\bibitem[27]{altlandZirnbauer97}
A.\ Altland and M.\ R.\ Zirnbauer, Phys. Rev. B \textbf{55}, 1142 (1997).

\bibitem[28]{volovikBOOKS}
G. E. Volovik, in 
\emph{The Universe in a Helium Droplet}, 
The International Series of Monographs on Physics Vol. 117 (Oxford University Press, New York, 2003); 
G. E. Volovik, in 
\emph{Exotic Properties of Superfluid 3He}, Series I N Modern Condensed Matter Physics Vol. 1 
(World Scientific, Singapore, 1992).

\bibitem[29]{roy2008} 
R.\ Roy, arXiv:0803.2868 (unpublished).

\bibitem[30]{zhangPRB08}
X.-L. Qi, T.\ Hughes, and S.-C.\ Zhang, Phys. Rev. B \textbf{78}, 195424 (2008).

\bibitem[31]{schnyderPRL09}
A.\ P.\ Schnyder, S.\ Ryu, and A.\ W.\ W.\ Ludwig, Phys. Rev. Lett. \textbf{102}, 196804 (2009).

\bibitem[32]{schnyderPRB10}
A.~P.~Schnyder, P.~M.~R.~Brydon, D.~Manske, and C.~Timm, Phys.~Rev.~B \textbf{82}, 184508 (2010).

\bibitem[33]{kaneMelea05a} 
C.\ Kane and E.\ Mele, Phys. Rev. Lett. \textbf{95}, 226801 (2005). 

\bibitem[34]{kaneMelea05b}
C.\ Kane and E.\ Mele, Phys. Rev. Lett. \textbf{95}, 146802 (2005). 

\bibitem[35]{fuKane07}
L.\ Fu and C.\ L.\ Kane, Phys.\ Rev.\ B \textbf{76} 045302 (2007).

\bibitem[36]{moorePRB07}
J.\ E.\ Moore and L.\ Balents, Phys.\ Rev.\ B \textbf{75} 121306 (2007).

\bibitem[37]{royPRB09}
R.\ Roy, Phys. Rev. B \textbf{79}, 195321 (2009).

\bibitem[38]{frigeriPRL04}
P.\ A.\ Frigeri, D.\ F.\ Agterberg, A.\ Koga, and M.\ Sigrist,
Phys.\ Rev.\ Lett.\ \textbf{92}, 097001 (2004).

\bibitem[39]{samokhinPRB08}
K.\ V.\ Samokhin and V.\ P.\ Mineev, Phys.\ Rev.\  B \textbf{77}, 104520 (2008).




\end{thebibliography}
\end{document}